\documentclass{emulateapj}
\usepackage{apjfonts}

\newcommand{\lsi}    {\object{LS~I~+61~303}}
\newcommand{\lsif}   {LS~I~+61~303}

\shorttitle{X-ray structure in \lsi}
\shortauthors{Paredes et al.}

\begin{document}

\title{{\it Chandra} Observations of the Gamma-ray Binary \lsi: Extended X-ray Structure?}

\author{
Josep M. Paredes,\altaffilmark{1}
Marc Rib\'o,\altaffilmark{1}
Valent\'{\i} Bosch-Ramon,\altaffilmark{2}\\
Jennifer R. West,\altaffilmark{3}
Yousaf M. Butt,\altaffilmark{3}
Diego F. Torres,\altaffilmark{4}
Josep Mart\'{\i}\altaffilmark{5}
}
\altaffiltext{1}{Departament d'Astronomia i Meteorologia, Facultat de F\'{\i}sica, Universitat de Barcelona, Mart\'{\i} i Franqu\`es 1, E-08028 Barcelona, Spain; jmparedes@ub.edu, mribo@am.ub.es.}
\altaffiltext{2}{Max Planck Institut f\"ur Kernphysik, D-69117 Heidelberg, Germany; vbosch@mpi-hd.mpg.de.}
\altaffiltext{3}{Harvard-Smithsonian Center for Astrophysics, 60 Garden Street, Cambridge, MA 02138; jennifer@head.cfa.harvard.edu, ybutt@head.cfa.harvard.edu.}
\altaffiltext{4}{Instituci\'o Catalana de Recerca i Estudis Avan\c cats (ICREA) \& Institut de Ci\`encies de l'Espai (IEEC-CSIC) Campus UAB, Fac. de Ci\`encies, Torre C5, parell, 2a planta, E-08193 Barcelona, Spain;
dtorres@aliga.ieec.uab.es.}
\altaffiltext{5}{Departamento de F\'{\i}sica, Escuela Polit\'ecnica Superior, Universidad de Ja\'en, Las Lagunillas s/n, Edificio A3, 23071 Ja\'en, Spain; jmarti@ujaen.es.}

\begin{abstract} 

We present a 50~ks observation of the gamma-ray binary \lsi\ carried out with
the ACIS-I array aboard the {\it Chandra} X-ray Observatory. This is the
highest resolution X-ray observation of the source conducted so far. Possible
evidence of an extended structure at a distance between 5\arcsec\ and
12\arcsec\ towards the North of \lsi\ have been found at a significance level
of 3.2$\sigma$. The asymmetry of the extended emission excludes an
interpretation in the context of a dust-scattered halo, suggesting an intrinsic
nature. On the other hand, while the obtained source flux, of $F_{\rm
0.3-10~keV}=7.1^{+1.8}_{-1.4}\times 10^{-12}$~ergs~cm$^{-2}$~s$^{-1}$, and
hydrogen column density, $N_{\rm H}=0.70\pm 0.06\times 10^{22}$~cm$^{-2}$, are
compatible with previous results, the photon index $\Gamma=1.25\pm 0.09$ is the
hardest ever found. In light of these new results, we briefly discuss the
physics behind the X-ray emission, the location of the emitter, and the
possible origin of the extended emission $\sim$0.1~pc away from \lsi.

\end{abstract}


\section{Introduction}

\lsi\ is a high mass X-ray binary associated with the galactic plane variable
radio source \object{GT~0236+610} \citep{gregory78} which shows periodic
non-thermal outbursts on average every $P_{\rm orb}$=26.4960$\pm$0.0028~d
\citep{taylor82,gregory02}. Optical spectroscopic observations show that the
system is composed of a rapidly rotating early type B0\,Ve star with a stable
equatorial decretion disk and mass loss, and a compact object with a mass
between 1 and 4 $M_{\odot}$ orbiting it every $\sim$26.5~d
\citep{hutchings81,casares05,grundstrom07}. Spectral line radio observations
give a distance of 2.0$\pm$0.2~kpc \citep{frail91}. \cite{massi01,massi04}
reported the discovery of an extended jet-like and apparently precessing radio
emitting structure at angular extensions of 10--50~milliarcseconds. Due to the
presence of (apparently relativistic) radio emitting jets, \lsi\ was proposed
to be a microquasar. However, recent VLBA images obtained during a full orbital
cycle show a rotating elongated morphology \citep{dhawan06}, which may be
consistent with a model based on the interaction between the relativistic wind
of a non-accreting pulsar and the wind of the stellar companion
\citep{dubus06}. The MAGIC Cerenkov telescope has recently detected \lsi\ at
very high energy gamma rays ($\gtrsim$100~GeV; \citealt{albert06}). \lsi\ and
\object{LS~5039} \citep{paredes00,aharonian05} share the quality of being the
only two known X-ray binaries that are also GeV emitters. Therefore, they can
be called gamma-ray binaries.

\lsi\ has been observed with different X-ray satellites: {\it Einstein}
\citep{bignami81}, {\it ROSAT} \citep{goldoni95,taylor96}, {\it ASCA}
\citep{leahy97}, {\it RXTE} \citep{harrison00}, {\it BeppoSAX}, {\it
XMM-Newton}, and {\it INTEGRAL} \citep{sidoli06,chernyakova06}. The spectrum
could always be fitted with an absorbed power-law with values in the ranges
$N_{\rm H}$=0.45--0.65$\times10^{22}$~cm$^{-2}$ and $\Gamma$=1.5--1.9. The
2--10~keV X-ray flux shows orbital variability \citep{paredes97}, with values
ranging from about 5 to 20$\times10^{-12}$~ergs~cm$^{-2}$~s$^{-1}$, the maximum
occurring at orbital phase $\sim$0.5 (assuming $T_0$=JD 2,443,366.775). No
spectral lines nor extended X-ray emission have ever been reported. Here we
present, for the first time, {\it Chandra} X-ray observations of \lsi, aimed at
detecting small-scale extended X-ray emission.

\section{{\it Chandra} observations and results}

We observed LS~I~+61~303 with {\it Chandra} using the standard ACIS-I setup in
VF mode and 3.24~s time frame during a total of 49.9~ks from 2006 April 7
22:30~UT to April 8 12:22~UT (MJD~53832.9375--53833.5153). This corresponds to
orbital phases 0.028--0.050, when a relatively low flux is expected, thus
minimizing possible pileup effects.

The {\it Chandra} Interactive Analysis of Observations software package
(CIAO~3.3.0.1) and the CALDB 3.2.2 version have been used to perform the
reduction of the level 1 event files as well as for the spectrum and lightcurve
subtraction. The reprocessed level 2 event files have been created to account
for readout effects. The position angle of the readout direction of the
ACIS-I chip was $-62$\degr\ (positive from North to East).

We have reduced the data using standard procedures given in the {\it Chandra}
analysis threads. The source region has been defined as a 20 pixel
($\sim$10\arcsec) radius circle around the source center. For the background,
we took a nearby circular region of 80 pixel ($\sim$40\arcsec) radius, away
from the source and the CCD junction. To avoid the presence of possible (very
faint) sources in the background, the tool {\it celldetect} was applied. The
{\it mkacisrmf} and {\it mkarf} tools were used to generate the RMF and ARF
files respectively. The tool {\it dmextract} was used to generate the source
and the background lightcurves and spectra. The spectrum was grouped with the
tool {\it dmgroup} to 30 counts per energy bin.

\lsi\ showed a moderate level of activity, with an average count rate of
0.15~counts~s$^{-1}$ (see Fig.~\ref{fig:lc}), and the data was affected by
pileup at the $\sim$14\% level (see below). The background was very low and
estimated to be about 1\% of the source count rate within the source region.
The position of the X-ray source, determined by using the CIAO package tool
{\it celldetect}, is $\alpha=02^{\rm h} 40^{\rm m} 31\fs63\pm0\fs02$,
$\delta=61\degr 13\arcmin 45\farcs70\pm0\farcs3$ (J2000). The errors, as in the
rest of this work, are 1$\sigma$. The X-ray position is consistent with the
radio and optical positions.

\subsection{Lightcurve}

\begin{figure}
\resizebox{1.0\hsize}{!}{\includegraphics[angle=0]{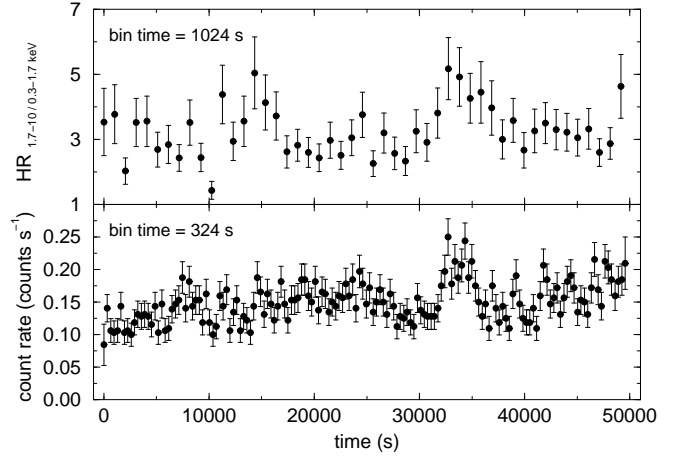}}
\caption{
{\it Bottom}: Background subtracted lightcurve of \lsif\ obtained with {\it
Chandra}. The binning time is 324~s and the error bars are at the 1$\sigma$
level. Time origin corresponds to 2006 April 7 at 22:33:17 UT (MJD~53832.9398),
and the observation length is 50~ks. Total count rate variations of a factor of
two on timescales of a few hours are clearly seen.
{\it Top}: Hardness ratio (HR) between 1.7--10 and 0.3--1.7~keV energy ranges.
The binning time has been increased to 1024~s to reduce the size of the error
bars. The HR is not significantly variable and it is poorly correlated with the
count rate (except during the miniflare).
\label{fig:lc}}
\end{figure}

The background subtracted lightcurve of \lsi\ in the 0.3--10~keV band is
plotted at the bottom of Fig.~\ref{fig:lc}. Each point represents 324~s of data
and the error bars represent the square root of the total number of counts in
each bin divided by 324~s. We have also computed the hardness ratio (HR) by
dividing the count rate in the 1.7--10~keV energy range by the corresponding
one in the 0.3--1.7~keV range. We show at the top of Fig.~\ref{fig:lc} the HR
as a function of time, but with a larger bin time of 1024~s to have relatively
small error bars. The count rate is moderately variable on timescales from
several minutes to hours, with an average of 0.144$\pm$0.029~counts~s$^{-1}$
($\chi^2_{\rm red}$=1.84). The count rate increases during the observation, and
a linear fit yields a 30\% increase at the end with respect to the beginning.
Superimposed on this trend is a miniflare 32~ks after the start of the
observation, when the count rate increases by a factor of $\sim$2 over a
timescale of $\sim$1000~s, with a total duration of $\sim$3~ks. This miniflare
is accompanied by an increase of the HR. This kind of moderate variability was
first observed with {\it ASCA}, which detected source variations of about a
50\% on timescales of half an hour \citep{harrison00}. Archival {\it BeppoSAX}
observations carried out in 1997 have also shown short-term variability
\citep{sidoli06}. It is interesting to note that small amplitude ($<$10\%)
radio flares were detected in \lsi\, with a recurrence period of about 5~ks
\citep{peracaula97}. The presence of hour timescale X-ray miniflares has also
been reported in the case of \object{LS~5039} \citep{bosch05}. A hardness vs.
intensity diagram for the whole {\it Chandra} observation of \lsi\ reveals a
positive trend, the source is harder when it is brighter, but with a low
significant correlation coefficient of $r$=0.44. This is in contrast to the
clear correlation found in previous {\it XMM} observations by \cite{sidoli06}.

\subsection{X-ray spectrum}

Since the HR is not significantly variable ($\chi^2_{\rm red}$=1.78), we
considered the whole data set for the spectral analysis. To perform the
spectral fits we restricted the energy range between 0.3 and 10~keV. A fit to
the data with a Bremsstrahlung model provided $\chi^2_{\rm red}$=1.6, and no
lines nor a multicolor blackbody were required to fit the data. An absorbed
power-law fit yielded $\chi^2_{\rm red}$=1.14, but a significant excess above
7~keV was apparent in the residuals. Due to the count rate and time frame used,
this is probably due to pileup. The addition of the pileup model implemented in
{\it Sherpa} \citep{davis01} to the absorbed power law yielded different sets
of possible spectral parameter values, all of them equally compatible from a
statistical point of view. Nevertheless, for some of them the derived fluxes
were a factor $\ga$5 higher than previous measurements at similar orbital
phases. We found that reasonable results were obtained when fixing the event
pileup fraction parameter $f$ to 0.95: $\chi^2_{\rm red}$=0.83, grade migration
parameter $\alpha=0.33\pm 0.09$, $N_{\rm H}=0.70\pm 0.06\times
10^{22}$~cm$^{-2}$, $\Gamma=1.25\pm 0.09$, pileup fraction 14\%, and $F_{\rm
0.3-10~keV}=7.1^{+1.8}_{-1.4}\times 10^{-12}$~ergs~cm$^{-2}$~s$^{-1}$. The
significance of adding a pileup component to the absorbed powerlaw fit was
estimated via an F-test: the probability that this improvement happens by
chance is 0.8\%.

To check the robustness of the results obtained with the spectral fitting with
the pileup model we have conducted two additional tests. First of all we have
obtained the spectrum of the readout streak, which is not affected by pileup,
after excluding the 5 pixel diameter region around the source. Since it only
contains 287~counts, we have fitted the spectrum using low count number
statistical weights. Fixing $N_{\rm H}$ to $0.7\times 10^{22}$~cm$^{-2}$ we
have obtained $\Gamma=1.21\pm 0.14$, which is consistent within errors to the
value given above. An additional test has been conducted by fitting the
spectrum of the source wings, which are not affected by pileup. We have built a
spectrum with data beyond a radius of 1.5\arcsec\ from the source center (i.e.,
excluding the inner $3\times3$ pixels, those most affected by pileup) and
fitted it with and without fixing $N_{\rm H}$, obtaining in both cases a photon
index harder than the one obtained with the pileup model. A similar test with
data beyond 2.0\arcsec\ has provided similar results. These tests clearly
suggest that the photon index value obtained with the pileup model is reliable,
and that the source was intrinsically hard during this observation. Moreover,
the estimate of the total source flux using the fitted flux of the source wings
and the fraction of the simulated PSF (see next section) counts contained in
the wings is compatible for $f$=0.95. This is not the case for the lower values
of $f$ that implied the much higher source fluxes discussed in the previous
paragraph. These results give us confidence on the results obtained with the
pileup model.

\begin{figure*}
\begin{center}
\resizebox{1.0\hsize}{!}{\includegraphics[angle=0]{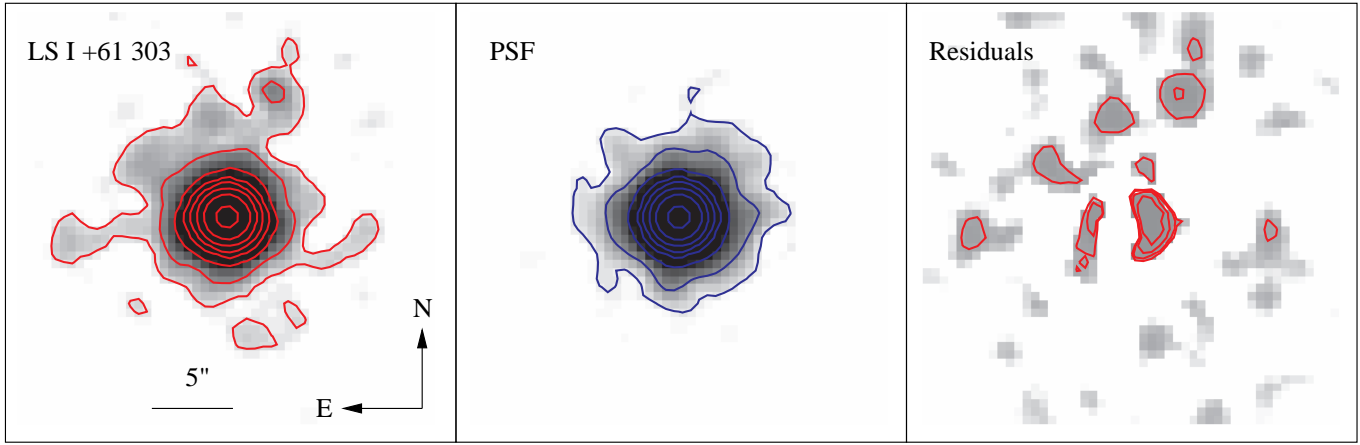}}
\caption{
{\it Left}: Image of \lsif\ obtained with {\it Chandra} in the 0.3--10 keV
energy range. The image size is $24\farcs5\times24\farcs5$ (the pixel size is
$0\farcs49\times0\farcs49$). A Gaussian kernel of three pixels in radius has
been used to smooth the image. The intensity scales logarithmically. The
contours correspond to seven logarithmic steps in the range
0.2--300~counts~pixel$^{-1}$. Extended X-ray emission is probably present at a
distance of 5--12\arcsec\ from the center.
{\it Middle}: Same as before but for the PSF.
{\it Right}: Same as before but for the residuals obtained after subtracting
the PSF image to the source one. The intensity scales here via a histogram
equalization. The extended emission at distances beyond 5\arcsec, in the
form of several spots towards the North-East, is apparent. The central
residuals at $\sim$2\arcsec\ radius occur within the core of the PSF, and no
conclusion can be drawn because of pileup.
\label{fig:images}}
\end{center}
\end{figure*}

Regarding the emission of \lsi\ itself, the flux we report here, $F_{\rm
0.3-10~keV}=7.1^{+1.8}_{-1.4}\times 10^{-12}$~ergs~cm$^{-2}$~s$^{-1}$, is compatible
within errors to that obtained during a short {\it XMM} observation also around
phase 0 \citep{chernyakova06}. However, the photon index $\Gamma=1.25\pm 0.09$
found in the {\it Chandra} data is the hardest ever found, only compatible with
any previous value at the $3\sigma$ level and harder, at a $\sim$5$\sigma$
level, than the {\it XMM} one at the same phase ($\Gamma=1.78\pm 0.04$). It is
outside the scope of this work to put forward a physical model for the spectral
state of the source. We will just point out that the X-ray photon index can be
harder than the canonical value of 1.5, which corresponds to a electron
population with power-law index of 2. This can occur for different reasons, for
example a Klein-Nishina dominated steady particle population (see
\citealt{derishev06} for an extended discussion). A very hard spectral
component (e.g. inverse Compton or Bremsstrahlung), produced by the lower
energy part of the particle spectrum with a high low-energy cutoff, may also
harden the synchrotron spectrum. The lack of simultaneous data at higher
energies does not allow us to distinguish between the two hypotheses.

We finally selected the data corresponding to the miniflare and obtained a
spectrum for it, which had poor statistics and was especially affected by
pileup. A fit to this spectrum yielded a photon index compatible within (large)
errors to that of the whole observation spectrum.

\subsection{Imaging}

A preliminary visual inspection of the source image reveals the possible
existence of extended emission. To better study this issue we have used {\it
ChaRT} and {\it MARX} to simulate the point spread function (PSF) accounting
for pileup effects. This PSF has the same counts as the source, which are
distributed in energy according to the source spectrum including pileup, being
spatially distributed in the image considering pileup as well\footnote{We note
that, while the pileup model included in {\it MARX} is valid for qualitative
predictions of the effects of pileup on the PSF, it has not been verified for
image reconstruction (section 7.3 of the {\it MARX} 4.0 Technical Manual).}. We
show in Fig.~\ref{fig:images} the images corresponding to the source, the PSF,
and the residuals obtained after subtracting the PSF from the source, all of
them smoothed with a Gaussian kernel of three pixels in radius to enhance the
faint emission. As can be seen in Fig.~\ref{fig:images}-left, there appears to
be extended X-ray emission around \lsi\ at distances between 5\arcsec\ and
12\arcsec\ from the source center. The subtraction of the nearly circular PSF,
shown in Fig.~\ref{fig:images}-middle, from the source image yields the
residuals shown in Fig.~\ref{fig:images}-right. There seems to be a region with
an excess of counts at several arcseconds from the image core and nearly
half-surrounding it. This excess region, already apparent in the source image,
presents some spots and extends mainly towards the North-East. 

\begin{figure}
\resizebox{1.0\hsize}{!}{\includegraphics[angle=0]{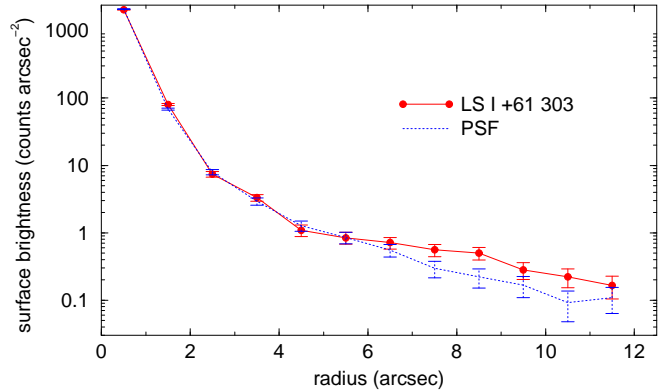}}
\caption{Surface brightness distribution of \lsif\ (red filled circles) compared to the PSF (dashed blue line). The excess above $\sim$5\arcsec\ is apparent in this radial profile, even if the extended emission is asymmetric.
\label{fig:radial}}
\end{figure}

To check the reliability of the 5--12\arcsec\ excess we have created radial
profiles of the surface brightness for the source and the simulated PSF with
the tools {\it dmextract} and {\it dmtcalc}. We show the radial profiles with
bins of 1\arcsec\ in Fig.~\ref{fig:radial}, where hints of excess are seen in
the region 5--12\arcsec\ away from the center. Moreover, we note that a
comparison with the unscattered point-like blazar source
\object{PKS~2155$-$304} located at high-galactic latitude (following the
procedure described in \citealt{gallo02}) provided an even clearer departure of
\lsi\ from a point-like source. Nevertheless, we prefer to show here the most
conservative results obtained with the PSF, simulated at the chip position and
with the observed spectrum of \lsi\ including pileup effects. Next we computed
the excess of source counts over the PSF in an annulus with inner and outer
radii of 5\arcsec\ and 12\farcs5, and obtained 58$\pm$18 counts (3.2$\sigma$,
or single trial probability of $1.4\times10^{-3}$). We have looked for excesses
in a circle with radius 20\arcsec\ centered on the source, without considering
the central 5\arcsec\ radius circle, thus yielding an area of
$\pi(20^2-5^2)$\arcsec$^2$. Dividing this by the area of the annulus where we
obtained the 3.2$\sigma$ excess we find that the number of trials is 2.9,
yielding a probability of $4.1\times10^{-3}$, which corresponds to 2.9$\sigma$.
However, since the extended emission appears to be asymmetric, we have computed
the excess dividing the 5--12\farcs5 annulus in four quadrants. Although there
is no excess towards the West and South directions, the excess is of about
1.5$\sigma$ to the East, and reaches 3.8$\sigma$ to the North (or single trial
probability of $1.4\times10^{-4}$). We note that this excess is not located in
the path of the readout streak. The number of trials is now 11.6, providing a
post-trial probability of $1.6\times10^{-3}$ or a 3.2$\sigma$ detection. The
excess in the Northern quadrant of the 5--12\farcs5 annulus, with a projected
distance of 0.05--0.12~pc from the center of \lsi, represents 0.5\% of the
source counts, implying (with $\Gamma$=1.25) a flux of
$\sim$4$\times10^{-14}$~ergs~cm$^{-2}$~s$^{-1}$ and $L_{\rm
X}\sim$2$\times10^{31}$~ergs~s$^{-1}$.

\section{Discussion}

The Chandra observations reported here show that \lsi\ presented moderate
fluxes of $F_{\rm 0.3-10~keV}=7.1^{+1.8}_{-1.4}\times
10^{-12}$~ergs~cm$^{-2}$~s$^{-1}$, as expected around orbital phase $\sim 0$
(\cite{sidoli06}). The source appeared to be moderately variable on timescales
of $\sim$1000~s. The hydrogen column density was $N_{\rm H}=0.70\pm 0.06\times
10^{22}$~cm$^{-2}$, similar to the values found in the past. The photon index
was $1.25\pm0.09$, harder than the previously measured values in the range
1.5--1.9. Finally, an excess between 5 and 12\farcs5 towards the North of \lsi\
has been detected at the 3.2$\sigma$ level (post-trial).

{\it Chandra} observations of X-ray binaries have provided the most detailed
X-ray images of these sources. Extended circular emission, produced by ISM dust
grains along the line of sight that scatter mainly the soft X-rays
\citep{predehl95}, has been found in most cases \citep{xiang05}. However, the
possible extended structure around \lsi\ we report here is asymmetric, suggesting its
intrinsic nature. Plausible explanations of this diffuse radiation could be
thermal Bremsstrahlung produced by hot and dense ISM close to the source,
perhaps heated and compressed by a large-scale outflow originated in \lsi, and
synchrotron or inverse Compton emission produced by non-thermal particles
accelerated at the termination region of the aforementioned outflow.  The first
possibility would require a medium with densities significantly higher than
those inferred from the ISM absorption towards \lsi. If non-thermal, the
extended emission would more likely be of synchrotron origin due to its shorter
timescales to radiate at a few keV than those of inverse Compton or
relativistic Bremsstrahlung. In any case, the asymmetric distribution of counts
could be related to the main direction of the outflow and/or to ISM
inhomogeneities. Regardless of the morphology, we can say that efficient energy
transport is probably taking place up to distances of at least $\sim$0.1~pc in
\lsi. However, only deeper high-resolution observations can help to better
detect the excess (currently at a 3.8$\sigma$ level, 3.2$\sigma$ post-trial)
and determine the precise morphology and underlying physics.

A final comment on the hydrogen column density appears necessary. The $N_{\rm
H}$ value inferred from spectral fits to all available X-ray observations is in
the range 0.45--0.70$\times 10^{22}$~cm$^{-2}$, while the $N_{\rm H}$ value of
the ISM inferred from UV/optical absorption is in the range 0.45--0.60$\times
10^{22}$~cm$^{-2}$. Therefore, the detected X-rays suffer small intrinsic
absorption, if any. Moreover no significant $N_{\rm H}$ orbital variations have
ever been detected (as pointed out previously by \citealt{leahy97} after {\it
ASCA} observations). In conclusion, the apparent lack of significant soft X-ray
absorption and the constancy of $N_{\rm H}$ along the orbit are difficult to be
explained considering the slow and dense material expelled from the companion
star. The X-ray emitter should be placed away from the binary system to produce
such observational results, in contrast with the expectations from classical
colliding wind and microquasar scenarios. This issue certainly requires further
investigation.\\

\acknowledgments

We thank Mike Nowak for helpful discussions on the analysis of the {\it Chandra} data.
J.M.P., M.R., V.B-R. and J.M. acknowledge support by DGI of the Spanish Ministerio de Educaci\'on y Ciencia (MEC) under grants AYA2004-07171-C02-01 and
AYA2004-07171-C02-02 and FEDER funds.
M.R. acknowledges financial support from MEC through a \emph{Juan de la Cierva} fellowship.
V.B-R. thanks the Max-Planck-Institut f\"ur Kernphysik for its support and kind
hospitality.
J.R.W. was supported by Cycle~7 Chandra Grant GO6-7025X.
D.F.T. has been supported by MEC under grant AYA-2006-00530, and the Guggenheim Foundation.
J.M. is also supported by PAI
of Junta de Andaluc\'{\i}a as research group FQM322.

\facility{{\it Chandra}}

\end{document}